\documentclass[twocolumn,showpacs,amsmath,amssymb,prl,floatfix]{revtex4-1}
\usepackage{graphicx}% Include figure files
\usepackage{bm}% bold math
\usepackage{amsmath,amsfonts}

\begin{document}

\title{Entropic enhancement of spatial correlations in a laser-driven Rydberg gas}

\author{C. Ates}
\author{I. Lesanovsky}

\affiliation{School of Physics and Astronomy, University of
Nottingham, Nottingham, NG7 2RD, UK}


\date{\today}

\begin{abstract}
In a laser-driven Rydberg gas the strong interaction between atoms excited to Rydberg states results in the formation of collective excitations. Atoms within a so-called blockade volume share a single Rydberg excitation, which is dynamically created and annihilated. For sufficiently long times this driven system approaches a steady state, which lends its properties from a maximum entropy state of a Tonks gas. Using this connection we show that spatial correlations between Rydberg atoms are controlled by the number of atoms contained within a blockade volume. For a small number the system favors a disordered arrangement of Rydberg atoms, whereas in the opposite limit Rydberg atoms tend to arrange in an increasingly ordered configuration. We argue that this is an entropic effect which is observable in current experiments.
\end{abstract}
%\pacs{}
\maketitle

Rydberg gases are currently intensely investigated as they constitute a versatile platform for probing strongly correlated phenomena in quantum many-body systems. The hallmark of many-body Rydberg physics is the excitation blockade, which inhibits the simultaneous laser excitation of two atoms within a given distance due to the strong interaction between Rydberg atoms \cite{Lukin01}. This blockade effect and the emerging exclusion volume critically determine the excitation dynamics of an ultracold Rydberg gas manifesting themselves in a strong suppression of excitation probabilities \cite{Tong04,Singer04} and non-Poissonian counting statistics \cite{CubelLiebisch05}.

Since the Rydberg blockade defines an exclusion volume around each excited atom it seems instructive to compare the situation in ultracold Rydberg gases with that encountered in classical hard objects models which are often viewed as paradigmatic  systems describing simple liquids. In these systems the binary interaction potential is modeled by an impenetrable, finite-size core associated with each particle. The thermodynamic properties of these simple exclusion volume models are dominated by entropy. Depending on the density of hard objects these systems may, in particular, exhibit density-density correlations over a considerable range or even undergo  crystallization in higher spatial dimensions (see \cite{Torquato10} and references therein).

Yet, the Rydberg blockade differs in two main aspects from the situation found in hard objects models. First, the effect is coherent involving superpositions of many quantum states, i.e., all atoms contained within the exclusion volume share a single delocalized Rydberg excitation. Second, the blockade is a \emph{dynamical} effect that is induced by the interplay between the resonant laser field and strong Rydberg-Rydberg interactions. Thus, excitations are continuously created and annihilated \cite{Raitzsch08,Reetz-Lamour08} leading to the formation of a gas in which the number of Rydberg atoms is \emph{not} conserved. This point appears to be critical for the emergence of long-range density-density correlations, as ordered phases in Rydberg gases have so far only been predicted for situations where states containing a fixed number of Rydberg atoms are deterministically excited  \cite{Pohl10,Schachenmayer11,Bijnen11}. For this \emph{dynamical crystallization} technique a tailored sequence of laser pulses is used to connect an initial state without Rydberg excitations to a state with an ordered  arrangement of Rydberg atoms.  However, spatial correlations that  extend well beyond the blockade radius and emerge \emph{spontaneously} (i.e. without specifically addressing the desired many-particle state) have, to the best of our knowledge, not been reported yet \cite{Robicheaux05,Wuester10,Schwarzkopf11}.

In this work we address the question whether pronounced density-density correlations can spontaneously develop
in strongly interacting Rydberg gases by analyzing their dynamics at long times. It has been demonstrated experimentally \cite{Heidemann07,Reetz-Lamour08,Loew09} and theoretically \cite{Weimer08,Olmos09-2,Lesanovsky10,Ates11-2} that observables like the number of excited atoms reach a steady state after some time.  Moreover, recent theoretical work on Rydberg lattice gas models has indicated that properties of the steady state can be understood from the perspective of thermodynamic ensembles \cite{Ates11-2}. Here, we use this insight and argue that  the equilibrium state of a Rydberg gas shares important properties with a thermal maximum entropy state of a Tonks gas \cite{Tonks36}. Connecting to this result and using purely entropic arguments we identify the figure of merit that controls the Rydberg density as the number of atoms contained within the blockade volume. The larger this quantity, the higher the density of excited atoms and the more enhanced the spatial correlations of the Rydberg atoms become.

\begin{figure}
\includegraphics[width=0.7\columnwidth]{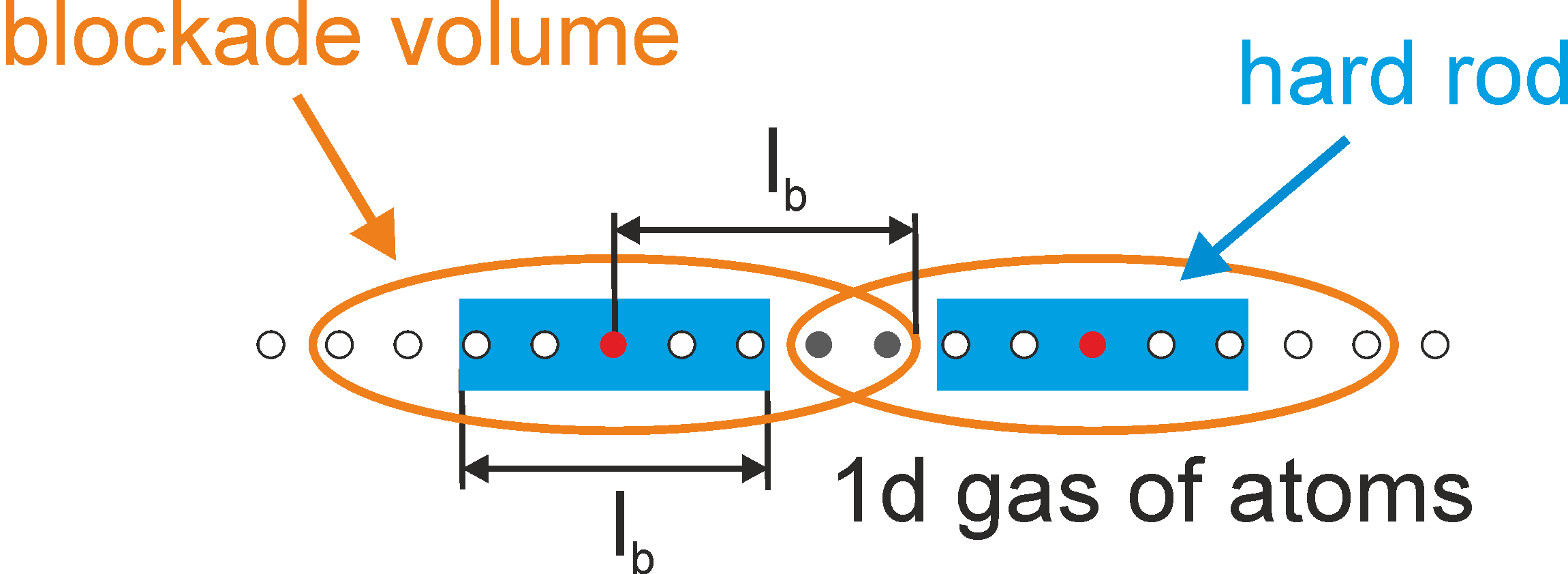}
\caption{
Excitation blockade and Tonks gas. Ground state atoms within a blockade radius $l_\mathrm{b}$ from a Rydberg atom (red solid circle) cannot be excited to a Rydberg state. This defines a blockade volume of size $2 l_{\text{b}}$ around each excited atom (orange ellipses), in which a single Rydberg excitation is shared between all particles. Certain sites may be located in the intersection of two blockade volumes (grey circles). In order to count the number of possible arrangements of Rydberg atoms it is convenient to map the system to a Tonks gas of non-overlapping hard rods (blue rectangles) of length $l_{\text{b}}$.
}
\label{fig:relaxation}
\end{figure}

We demonstrate this for a one-dimensional system. This has the advantage that a number of analytical results can be derived and numerically exact simulations of the quantum dynamics of $50-100$ particles are feasible.
We consider a gas with atomic line density $\sigma$ and large but \emph{fixed} length $L$. For the sake of simplicity (the physical mechanism discussed here does not depend on that) we consider atoms that are arranged on a regular lattice with spacing $\sigma^{-1}$ so that the number of lattice sites is $\sigma L$. The density of Rydberg atoms is denoted as $\rho$. Due to the blockade the maximum possible Rydberg density is limited to $\rho_\mathrm{max} \approx 1/l_\mathrm{b}$, where $l_{\text{b}}$ is the blockade radius (cf. Fig.\ \ref{fig:relaxation}), which we will also assume to be \emph{fixed} in the following. The internal structure of the atoms is modeled by two states: $\left|\downarrow\right>$ is the ground state and $\left|\uparrow\right>$ is the Rydberg state. Both states are resonantly coupled by a laser of Rabi frequency $\Omega$. The interaction between two excited atoms separated by a distance $r$ is given by the van-der-Waals potential $V(r)=C_6/r^6$ with dispersion coefficient $C_6$. The Hamiltonian of the system is
\begin{eqnarray}
  H_\mathrm{Ryd}=\Omega \sum^{\sigma L}_{k=1} \sigma_x^k + C_6\, \sigma^6 \sum^{\sigma L}_{k\neq m} \frac{n_k n_m}{|k-m|^6}
\end{eqnarray}
with $n_k=(\sigma^k_z+1)/2$ and the Pauli matrices $\sigma_\alpha$.
For sufficiently large interaction the simultaneous excitation of nearby Rydberg atoms is strongly suppressed. The corresponding exclusion or blockade radius $l_\mathrm{b}$ is calculated by setting the collective laser coupling of atoms within a blockade radius $\Omega\sqrt{\sigma\, l_\mathrm{b}}$ equal to the interaction energy $V(r)$ of two Rydberg atoms at a distance $l_\mathrm{b}$ \cite{Loew09} yielding  $l_\mathrm{b}=[C_6/(\Omega\sqrt{\sigma})]^{2/13}$.
This length is in fact not sharp and there is a finite though small probability that two atoms at a distance closer than $l_\mathrm{b}$ are simultaneously excited. However, in experiments and in theoretical calculations it has been shown that the assumption of a sharp blockade radius accurately captures the physics of in an interacting Rydberg gas \cite{Weimer08}. Furthermore, we also neglect all residual interactions beyond $l_\mathrm{b}$ due to the tail of the van-der-Waals potential as they are not essential to obtain a qualitative picture. Later on, we will further elucidate the consequences of this spatial cut-off. Within these approximations the Hamiltonian can be transformed to the form \cite{Lesanovsky11,Ji11,Ates11-2}
\begin{eqnarray}
  H=\Omega\sum^{\sigma L}_{k=1} \left[\prod_{m=k-\sigma l_\mathrm{b}, m\neq k}^{k+\sigma l_\mathrm{b}} (1-n_m)\right]\,\sigma_x^k\label{eq:hamiltonian}
\end{eqnarray}
where the exclusion of Rydberg atoms is made manifest by the projection operator formed by the term inside the square brackets. This projector probes whether there are excitations within the exclusion length from a given atom. If so, it yields zero while it is one in the opposite case.

The Hilbert space of the system consists of a number of subspaces which are not coupled by Hamiltonian (\ref{eq:hamiltonian}). We are interested here in the experimentally relevant subspace which contains the configuration $\left|0\right>=\left|\downarrow\downarrow\downarrow...\downarrow\right>$, i.e. the state without any Rydberg atom, and all configurations in which there are no two excitations within a blockade volume (orange ellipses in Fig.\ \ref{fig:relaxation}). All these classical configurations span the portion $\mathcal{H}_{\text{dyn}}$ of the Hilbert space in which the system evolves. In order to count the number of these configurations - which equals the dimension of $\mathcal{H}_{\text{dyn}}$ - it is convenient to map the physical situation, where blockade volumes may overlap to a system of non-overlapping hard rods (blue rectangles in Fig.\ \ref{fig:relaxation}). These hard rods have a diameter $l_{\text{b}}$ and occupy $\sigma l_\mathrm{b}+1$ lattice sites. Counting the number of ways for placing these hard rods reduces to the combinatorics of a lattice Tonks gas \cite{Chowdhury00}. This immediately yields the number of configurations containing $N$ Rydberg atoms
\begin{eqnarray}
  \Xi_L(N)=\frac{[\sigma L-N\sigma l_\mathrm{b}]!}{N!\, [\sigma L - N (\sigma l_\mathrm{b}+1)]!}\label{eq:entropy}
\end{eqnarray}
and the Hilbert space dimension $\mathrm{dim}\, \mathcal{H}_{\text{dyn}} =\sum^{\sigma L/(\sigma l_\mathrm{b}+1)}_{N=0}\Xi_L(N)$.

In one and two dimensions it has been shown that Hamiltonians of the form (\ref{eq:hamiltonian}) induce quantum dynamics that lead to a steady state in which all configurations spanning $\mathcal{H}_{\text{dyn}}$ are populated with equal probability \cite{Lesanovsky10,Ates11-2}. The steady state is, therefore, given by a a microcanonical maximum entropy state in the sense that expectation values of observables can be calculated from a microcanonical partition function. The partition sum for the system under study here is in fact given by Eq.\ (\ref{eq:entropy}). This is the connection of the long-time dynamics of a strongly interacting Rydberg gas with the thermal state of classical hard objects models.

\begin{figure}
\includegraphics[width=.8\columnwidth]{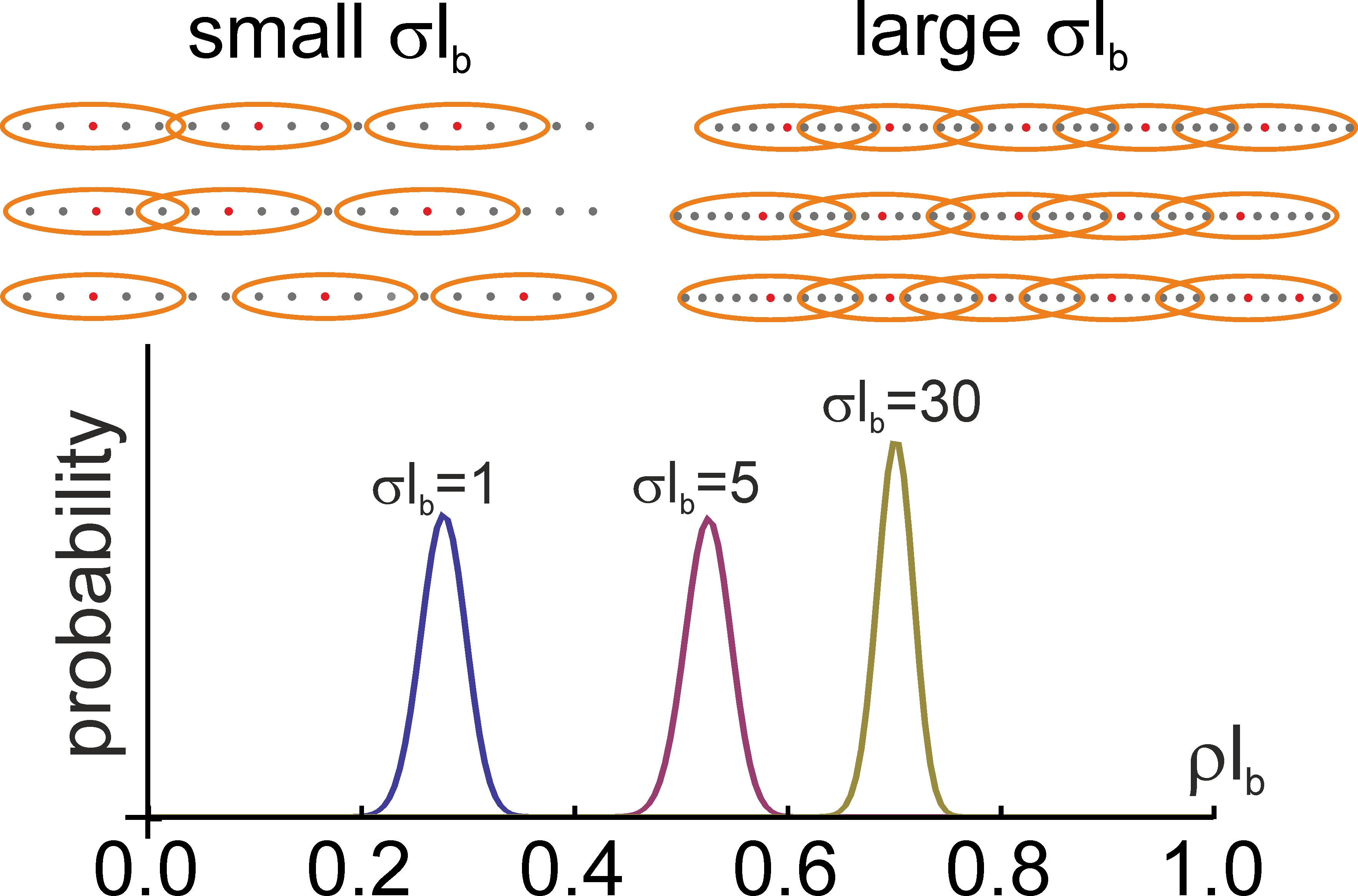}
\caption{Probability for finding configurations with a given density $\rho$
of Rydberg atoms for different values of $\sigma\,l_\mathrm{b}$ in a system of length $L=200\times l_\mathrm{b}$. For a small number of ground state atoms within a blockade radius ($\sigma l_\mathrm{b}$) the function peaks at small densities $\rho\ll l^{-1}_\mathrm{b}$ (see top left panel for three typical configurations). When $\sigma l_\mathrm{b}$ grows the relative weight of configurations with a large Rydberg density $\rho\sim l^{-1}_\mathrm{b}$ increases dramatically. Here a huge number of configurations with nearly crystalline arrangement exist (see top right panel for three examples). Hence, when all configurations are populated with equal probability, a state with high density and enhanced spatial correlations is entropically favored.}
\label{fig:entropy}
\end{figure}

For our one-dimensional setup we can use this connection to compute the steady state properties analytically. The number of Rydberg atoms $\langle N \rangle$ in equilibrium can be obtained from the distribution function $\Xi_L(N)$ using $\left<N\right>=(\mathrm{dim}\, \mathcal{H}_{\text{dyn}})^{-1} \sum^{\sigma L/(\sigma l_\mathrm{b}+1)}_{N=0}N\,\Xi_L(N)$. In the large volume limit ($L \gg l_{\text{b}}$) this function is strongly peaked. Thus, the mean number of Rydberg excitations in the steady state is given by the position of this peak which can be calculated using $\partial_N \log \Xi_L(N)|_{N=\langle N \rangle}=0$. Assuming further that the number of atoms within each blockade volume is large ($\sigma l_\mathrm{b}\gg 1$) we obtain the relationship
\begin{eqnarray}
  \left[\frac{\sigma}{\rho}-\sigma l_\mathrm{b}\right]\log\left(\frac{\sigma}{\rho}-\sigma l_\mathrm{b}\right)=\sigma l_\mathrm{b}
\end{eqnarray}
with the line density of Rydberg atoms $\rho=\langle N \rangle/L$. Using the Lambert $W$-function defined by $W(x)\,e^{W(x)}=x$ this can explicitly be computed:
\begin{eqnarray}
  \rho=\frac{1}{l_\mathrm{b}}\frac{W(\sigma l_\mathrm{b})}{1+W(\sigma l_\mathrm{b})}.\label{eq:rydberg_density}
\end{eqnarray}
An analysis of this expression shows that for fixed length $L$ the density of Rydberg atoms in the steady state grows as the total atomic density and, therefore, the number $\sigma l_\mathrm{b}$ of atoms per blockade radius is increased. This behavior is shown in Fig.\ \ref{fig:entropy}, where the graph depicts the probability for finding configurations with a given density $\rho$ of Rydberg atoms for different values of $\sigma\,l_\mathrm{b}$ in a system of large but fixed size $L$ and fixed $l_\mathrm{b}\ll L$.

In addition to the enhancement of the Rydberg density, the number fluctuations relative to the mean number of excited atoms decrease with increasing number of particles in the blockade volume. To characterize this we determine the Mandel Q-parameter: $Q=\frac{\langle N^2 \rangle-\langle N \rangle^2}{\langle N \rangle}-1$. In the vicinity of its maximum the function (\ref{eq:entropy}) can be approximated by a Gaussian. The fluctuations in the number of Rydberg atoms are then given by $\langle N^2 \rangle-\langle N \rangle^2=-1/\left(\partial^2_N \log \Xi_L(N)|_{N=\langle N \rangle}\right)$, yielding
\begin{eqnarray}
  Q=(\rho l_\mathrm{b})^2-2\rho l_\mathrm{b}=\left[1+W(\sigma l_\mathrm{b})\right]^{-2}-1.\label{eq:q-parameter}
\end{eqnarray}
The $Q$-parameter decreases with increasing $\sigma l_\mathrm{b}$, i.e.\ the distribution function depicted in Fig. \ref{fig:entropy} becomes increasingly sub-Poissonian. Since the blockade radius $l_\mathrm{b}$ is held constant and
the density of Rydberg atoms increases this behavior of the fluctuations indicates an \emph{enhancement of spatial correlations between the excited atoms}.

The fact that pronounced spatial correlations emerge when just the total atomic density $\sigma$ is increased, is a direct result of the counting of all configurations which led to Eq.\ (\ref{eq:entropy}). This is a somewhat counterintuitive finding. One might expect that the maximum of the entropy function ($\propto \log \Xi_L(N)$) lies at low Rydberg densities irrespective of the lattice spacing, since there is a vast number of possibilities for exciting a relatively small number of Rydberg atoms on the lattice. However, it is actually also possible to gain entropy with nearly ordered configurations. To illustrate this, suppose we have a configuration in which $n$ Rydberg atoms are already present. Then, the number of possibilities to place an additional excitation is given by the number of sites that are not enclosed by any blockade volume. Hence, for closest packing of the $n$ Rydberg atoms the number of available sites is maximized. This is because $(n-1) \sigma l_{\text{b}}$ blockaded sites are shared among the already excited atoms (grey sites in Fig.\ \ref{fig:relaxation} showing the case of non-maximum overlap of blockade volumes). Clearly, this dense packing of $n$ Rydberg atoms increases the entropy for placing another one. This effect becomes more pronounced the larger the number of atoms contained in a blockade volume. Eventually this outweighs the tendency of the maximum entropy state to be formed by low density configurations. Thus, in summary, for small values of $\sigma l_{\text{b}}$ the steady state mostly contains low density configurations  (cf.\ top left panel in Fig.\ \ref{fig:entropy}) while in the opposite limit large Rydberg densities are favored (cf.\ top right panel in Fig.\ \ref{fig:entropy}). The latter implies enhanced spatial correlations.
\begin{figure}
\includegraphics[width=\columnwidth]{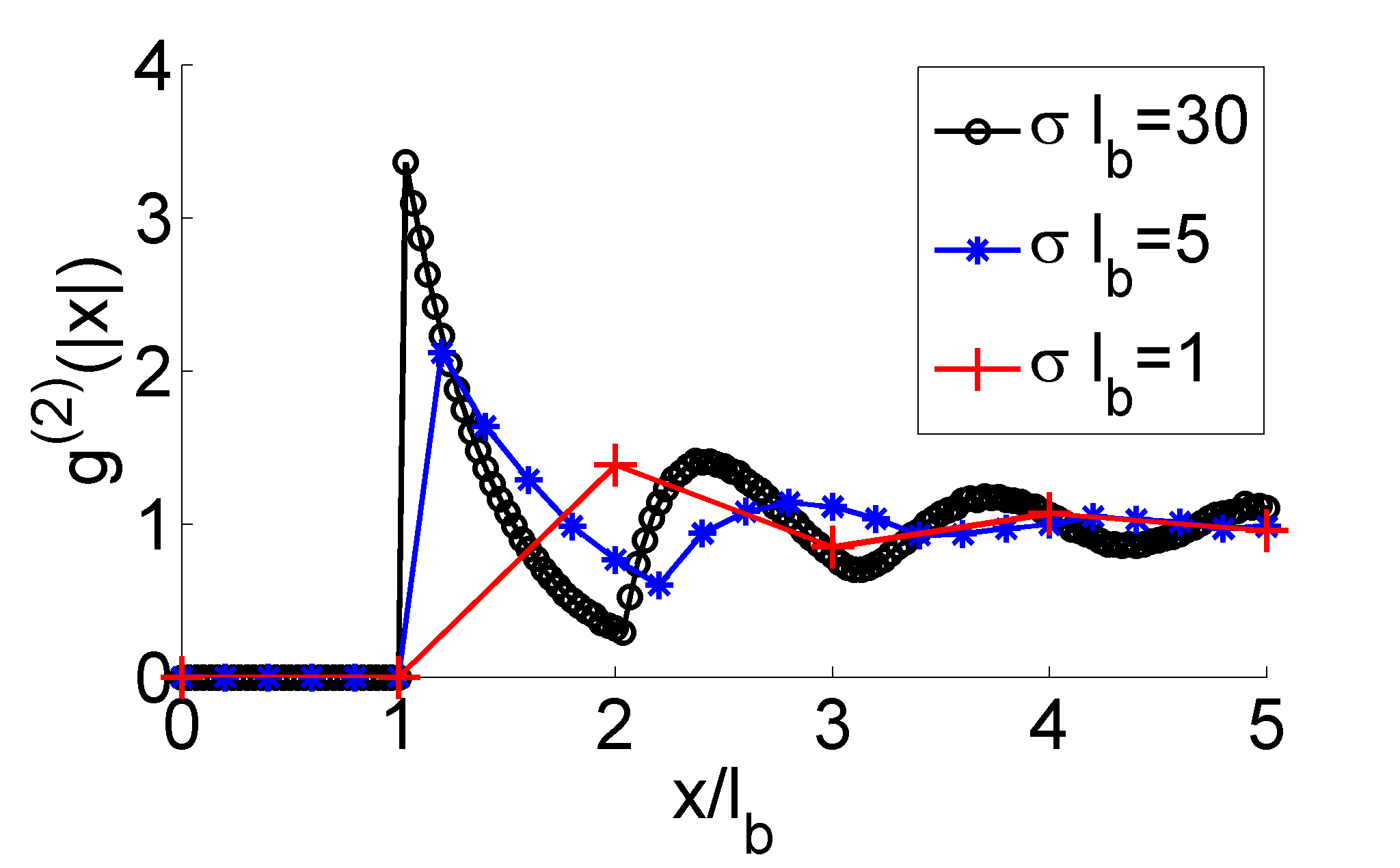}
\caption{Density-density correlation function obtained from a Monte-Carlo simulation for a sample of length $L=10\times l_\mathrm{b}$ (periodic boundary conditions) and atomic densities corresponding to $ \sigma l_\mathrm{b}=\{1, 5, 30\}$. The larger the number of ground state atoms per blockade radius $\sigma l_\mathrm{b}$ the more pronounced become the oscillations of the spatial correlations.}
\label{fig:correlations}
\end{figure}

To characterize these correlations in more detail, let us study the Rydberg-Rydberg correlation function $g^{(2)}(|x|)=\langle N(0) N(x)\rangle/\langle N \rangle^2$. To calculate $g^{(2)}(|x|)$ we use a classical Monte-Carlo method that generates a representative sample of classical configurations. In Fig.\ \ref{fig:correlations} we show data for a sample of length $L=10\times l_\mathrm{b}$ with periodic boundaries and for various values of the atomic line density: $\sigma l_\mathrm{b}=\{1, 5, 30\}$. Note that because of the underlying lattice, $x$ is only defined at positions where $\sigma\, x$ assumes integer values. We find that for an increasing number of ground state atoms per blockade radius the oscillations in the correlation function become more pronounced. This is consistent with the previously discussed decrease of the Q-parameter (\ref{eq:q-parameter}). For $\sigma l_\mathrm{b}\gg 1$ the discreteness of the ground state gas can be neglected and the correlation function is well approximated by that of a Tonks gas \cite{Salsburg53}
\begin{align}
  g_\mathrm{Tonks}^{(2)}(|x|)=\frac{e^{-W(\sigma l_\mathrm{b})\,|x|/l_\mathrm{b}}}{\rho\,l_\mathrm{b}}
  \sum^\infty_{k=1}\frac{\theta\left(x_k\right) (\sigma l_\mathrm{b}\,x_k)^k}{x_k\,(k-1)!}
\end{align}
with $x_k=|x|/l_\mathrm{b}-k$ and the unit step function $\theta\left(x\right)$. From this we see that the first peak decays exponentially with decay constant $[l_\mathrm{b}\,W(\sigma l_\mathrm{b})]^{-1}$ and
the height of the peak is given by $g_\mathrm{Tonks}^{(2)}(l_\mathrm{b})=W(\sigma l_\mathrm{b})+1$. This agrees with our simulation results.

\begin{figure}
\includegraphics[width=\columnwidth]{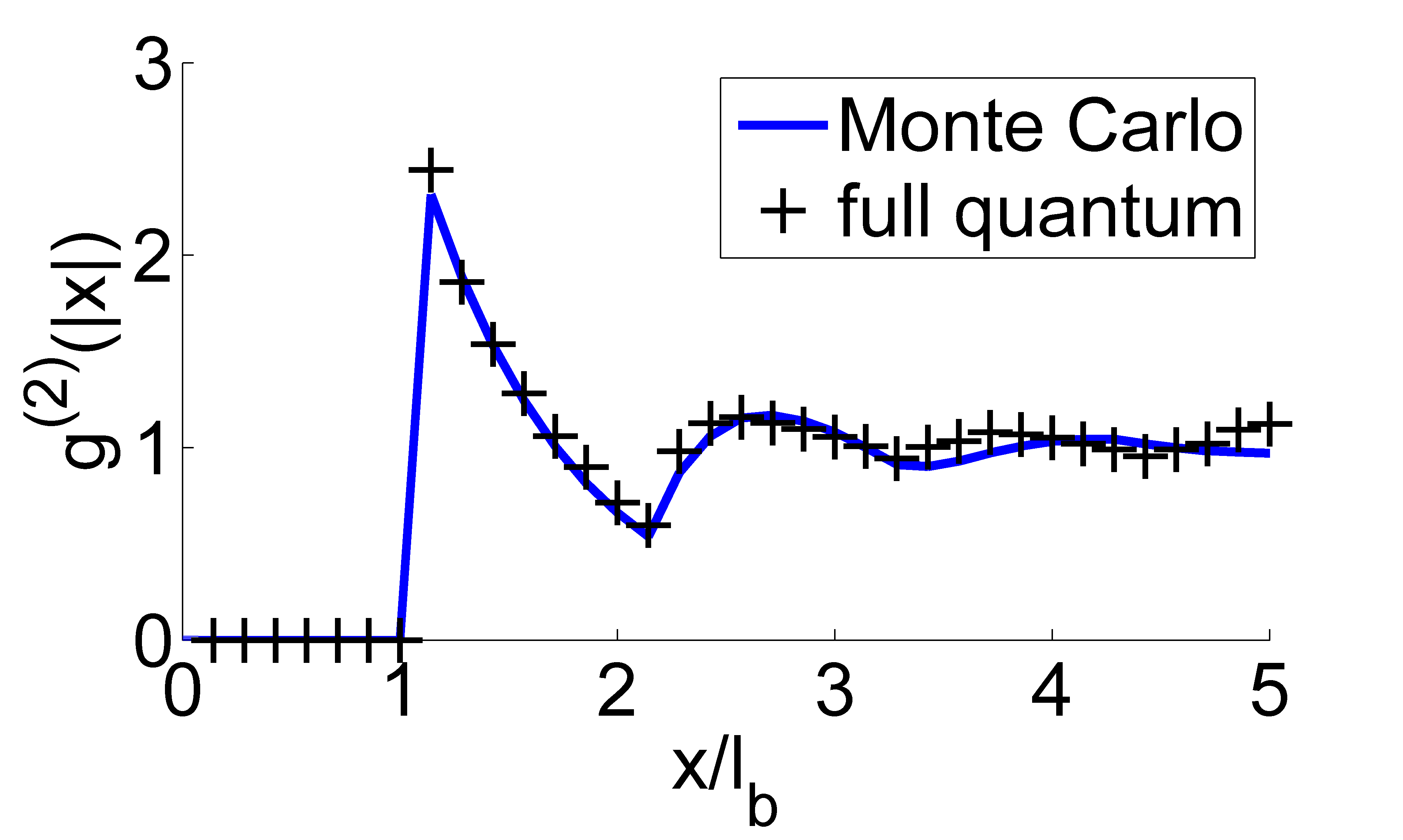}
\caption{Density-density correlation function for a sample of length $L=10\times l_\mathrm{b}$ (periodic boundary conditions) and an atomic density of $\sigma l_\mathrm{b}=7$. The crosses are obtained from a time-averaged quantum calculation using Hamiltonian (\ref{eq:hamiltonian}) and the empty state $\left|0\right>$ as an initial state. The solid curve interpolates data obtained from a (classical) Monte-Carlo simulation for the same parameters.}
\label{fig:quantum}
\end{figure}
Let us finally show that the steady state of the quantum evolution is indeed well captured by the statistical treatment and that we can indeed expect entropy induced spatial correlations to emerge in the coherently evolving quantum system. We have performed a numerical propagation of the initial state $\left|0\right>$ under the Hamiltonian (\ref{eq:hamiltonian}). Due to the exponential scaling of the Hilbert space dimension this can only be achieved within a rather limited parameter range. We have chosen $\sigma l_\mathrm{b}=7$, a system size of $10\times l_\mathrm{b}$ and a final propagation time of $\Omega t = 150$. Here the system is well within the steady state regime. In Fig.\ \ref{fig:quantum} we show the density-density correlation function averaged over the time-interval $130\leq \Omega t\leq 150$. The averaging removes small temporal fluctuations that are due to the finite size of the system (see Refs. \cite{Olmos09-2,Lesanovsky10,Ates11-2}). The agreement of the data (crosses) to the results from the classical Monte-Carlo simulation (solid line) is good. The first two peaks are almost perfectly reproduced, with the first one showing the expected exponential decay and the correct height. The agreement becomes less accurate with increasing $|x|$ but pronounced oscillation in the spatial correlations are nevertheless visible also here. The discrepancy with respect to the Monte-Carlo result is due to finite size effects and the particular choice of the initial state as discussed in Refs. \cite{Olmos09-2,Ates11-2}.

One-dimensional Rydberg gases as discussed here can be realized in elongated atomic clouds with a transverse extension much smaller than $l_\mathrm{b}$, see e.g. \cite{Loew09,Viteau11}. Since $l_\mathrm{b}$ is typically on the order of $5 \dots 10\, \mu\text{m}$, currently achievable densities would allow values of $\sigma l_{\text{b}} = 30$ or even larger. Yet, as compared to an experimental realization, the simulations presented here were undertaken for idealized conditions, that is, (i) an equidistant uniform distribution of atoms and (ii) no influence of the tail of the interaction potential. Both points can, in principle, be realized with available experimental technology.  The first condition can be approximately achieved in an optical lattice \cite{Viteau11}, the second by using microwave dressing techniques to ''chop off'' the tail of the interaction potential \cite{Buechler07,Micheli07}. But, even for experiments in a continuous gas and without tailoring the Rydberg-Rydberg potential our analytical results remain qualitatively valid.

In the disordered gas the atomic positions will vary from shot to shot, i.e., the number of atoms per blockade volume fluctuates which will lead to a higher $Q$ parameter than predicted by Eq.\ (\ref{eq:q-parameter}). The significance of these fluctuations, however, decreases with increasing atomic density $\sigma$, so that for sufficiently large densities our lattice gas approach becomes more and more accurate. The tail of the van-der-Waals potential will start to play a role for Rydberg densities close to $1/l_{\text{b}}$. In this regime, the concept of a fixed blockade radius is no longer accurate, as atoms that are not enclosed by any ''blockade volume'' may still be blockaded by the addition of the potential tails of two nearby Rydberg atoms. This will lead to a saturation the maximum achievable enhancement of Rydberg-Rydberg correlations \cite{Gaerttner12}. Due to the short-range character of the van-der-Waals potential this saturation effect will, however, happen at comparatively large Rydberg densities. Most importantly, the interesting effect of entropically enhanced spatial correlations of Rydberg atoms is already significant at relatively low Rydberg densities and will, thus, be experimentally observable.

In one dimension correlations always decay exponentially. In the future it will therefore be interesting to study the discussed system in higher dimensions as here long range order can emerge. This is difficult to treat theoretically but could be observable with current experiments together with appropriate imaging techniques \cite{Olmos11,Schwarzkopf11,Gunter12}.

\begin{acknowledgements}
\emph{Acknowledgements --- }
We acknowledge fruitful discussions with J. Evers, M. G\"arttner and J. P. Garrahan. This work was funded in part by EPSRC Grant no.  EP/I017828/1 and Leverhulme Trust grant no. F/00114/BG. C.A. acknowledges support through a Feodor-Lynen Fellowship of the Alexander von Humboldt Foundation.
\end{acknowledgements}

%\bibliographystyle{apsrev4-1}
%\bibliography{bib_Ryd}{}

\begin{thebibliography}{10}%
\makeatletter
\providecommand \@ifxundefined [1]{%
 \ifx #1\undefined \expandafter \@firstoftwo
 \else \expandafter \@secondoftwo
\fi
}%
\providecommand \@ifnum [1]{%
 \ifnum #1\expandafter \@firstoftwo
 \else \expandafter \@secondoftwo
\fi
}%
\providecommand \enquote [1]{``#1''}%
\providecommand \bibnamefont  [1]{#1}%
\providecommand \bibfnamefont [1]{#1}%
\providecommand \citenamefont [1]{#1}%
\providecommand\href[0]{\@sanitize\@href}%
\providecommand\@href[1]{\endgroup\@@startlink{#1}\endgroup\@@href}%
\providecommand\@@href[1]{#1\@@endlink}%
\providecommand \@sanitize [0]{\begingroup\catcode`\&12\catcode`\#12\relax}%
\@ifxundefined \pdfoutput {\@firstoftwo}{%
 \@ifnum{\z@=\pdfoutput}{\@firstoftwo}{\@secondoftwo}%
}{%
 \providecommand\@@startlink[1]{\leavevmode}%
 \providecommand\@@endlink[0]{}%
}{%
 \providecommand\@@startlink[1]{%
  \leavevmode
  \pdfstartlink
   attr{/Border[0 0 1 ]/H/I/C[0 1 1]}%
   user{/Subtype/Link/A<</Type/Action/S/URI/URI(#1)>>}%
  \relax
 }%
 \providecommand\@@endlink[0]{\pdfendlink}%
}%
\providecommand \url  [0]{\begingroup\@sanitize \@url }%
\providecommand \@url [1]{\endgroup\@href {#1}{\urlprefix}}%
\providecommand \urlprefix [0]{URL }%
\providecommand \Eprint[0]{\href }%
\@ifxundefined \urlstyle {%
  \providecommand \doi [1]{doi:\discretionary{}{}{}#1}%
}{%
  \providecommand \doi [0]{doi:\discretionary{}{}{}\begingroup
  \urlstyle{rm}\Url }%
}%
\providecommand \doibase [0]{http://dx.doi.org/}%
\providecommand \Doi[1]{\href{\doibase#1}}%
\providecommand \bibAnnote [3]{%
  \BibitemShut{#1}%
  \begin{quotation}\noindent
    \textsc{Key:}\ #2\\\textsc{Annotation:}\ #3%
  \end{quotation}%
}%
\providecommand \bibAnnoteFile [2]{%
  \IfFileExists{#2}{\bibAnnote {#1} {#2} {\input{#2}}}{}%
}%
\providecommand \typeout [0]{\immediate \write \m@ne }%
\providecommand \selectlanguage [0]{\@gobble}%
\providecommand \bibinfo [0]{\@secondoftwo}%
\providecommand \bibfield [0]{\@secondoftwo}%
\providecommand \translation [1]{[#1]}%
\providecommand \BibitemOpen[0]{}%
\providecommand \bibitemStop [0]{}%
\providecommand \bibitemNoStop [0]{.\EOS\space}%
\providecommand \EOS [0]{\spacefactor3000\relax}%
\providecommand \BibitemShut [1]{\csname bibitem#1\endcsname}%
%</preamble>
\bibitem{Lukin01}%
  \BibitemOpen
  \bibfield{author}{%
  \bibinfo {author} {\bibfnamefont{M.~D.}\ \bibnamefont{Lukin}}, \bibinfo
  {author} {\bibfnamefont{M.}~\bibnamefont{Fleischhauer}}, \bibinfo {author}
  {\bibfnamefont{R.}~\bibnamefont{C\^{o}t\'{e}}}, \bibinfo {author}
  {\bibfnamefont{L.~M.}\ \bibnamefont{Duan}}, \bibinfo {author}
  {\bibfnamefont{D.}~\bibnamefont{Jaksch}}, \bibinfo {author}
  {\bibfnamefont{J.~I.}\ \bibnamefont{Cirac}},\ and\ \bibinfo {author}
  {\bibfnamefont{P.}~\bibnamefont{Zoller}},\ }%
  \bibfield{journal}{%
  \bibinfo {journal} {Phys. Rev. Lett.}\ }%
  \textbf{\bibinfo {volume} {87}},\ \bibinfo {pages} {037901} (\bibinfo {year}
  {2001})%
  \bibAnnoteFile{NoStop}{Lukin01}%
\bibitem{Tong04}%
  \BibitemOpen
  \bibfield{author}{%
  \bibinfo {author} {\bibfnamefont{D.}~\bibnamefont{Tong}}, \bibinfo {author}
  {\bibfnamefont{S.~M.}\ \bibnamefont{Farooqi}}, \bibinfo {author}
  {\bibfnamefont{J.}~\bibnamefont{Stanojevic}}, \bibinfo {author}
  {\bibfnamefont{S.}~\bibnamefont{Krishnan}}, \bibinfo {author}
  {\bibfnamefont{Y.~P.}\ \bibnamefont{Zhang}}, \bibinfo {author}
  {\bibfnamefont{R.}~\bibnamefont{C\^ot\'e}}, \bibinfo {author}
  {\bibfnamefont{E.~E.}\ \bibnamefont{Eyler}},\ and\ \bibinfo {author}
  {\bibfnamefont{P.~L.}\ \bibnamefont{Gould}},\ }%
  \bibfield{journal}{%
  \Doi{10.1103/PhysRevLett.93.063001}{\bibinfo {journal} {Phys. Rev. Lett.}}\
  }%
  \textbf{\bibinfo {volume} {93}},\ \bibinfo {pages} {063001} (\bibinfo {year}
  {2004})%
  \bibAnnoteFile{NoStop}{Tong04}%
\bibitem{Singer04}%
  \BibitemOpen
  \bibfield{author}{%
  \bibinfo {author} {\bibfnamefont{K.}~\bibnamefont{Singer}}, \bibinfo {author}
  {\bibfnamefont{M.}~\bibnamefont{Reetz-Lamour}}, \bibinfo {author}
  {\bibfnamefont{T.}~\bibnamefont{Amthor}}, \bibinfo {author}
  {\bibfnamefont{L.~G.}\ \bibnamefont{Marcassa}},\ and\ \bibinfo {author}
  {\bibfnamefont{M.}~\bibnamefont{Weidem\"uller}},\ }%
  \bibfield{journal}{%
  \Doi{10.1103/PhysRevLett.93.163001}{\bibinfo {journal} {Phys. Rev. Lett.}}\
  }%
  \textbf{\bibinfo {volume} {93}},\ \bibinfo {pages} {163001} (\bibinfo {year}
  {2004})%
  \bibAnnoteFile{NoStop}{Singer04}%
\bibitem{CubelLiebisch05}%
  \BibitemOpen
  \bibfield{author}{%
  \bibinfo {author} {\bibfnamefont{T.~C.}\ \bibnamefont{Liebisch}}, \bibinfo
  {author} {\bibfnamefont{A.}~\bibnamefont{Reinhard}}, \bibinfo {author}
  {\bibfnamefont{P.~R.}\ \bibnamefont{Berman}},\ and\ \bibinfo {author}
  {\bibfnamefont{G.}~\bibnamefont{Raithel}},\ }%
  \bibfield{journal}{%
  \Doi{10.1103/PhysRevLett.95.253002}{\bibinfo {journal} {Phys. Rev. Lett.}}\
  }%
  \textbf{\bibinfo {volume} {95}},\ \bibinfo {pages} {253002} (\bibinfo {year}
  {2005})%
  \bibAnnoteFile{NoStop}{CubelLiebisch05}%
\bibitem{Torquato10}%
  \BibitemOpen
  \bibfield{author}{%
  \bibinfo {author} {\bibfnamefont{S.}~\bibnamefont{Torquato}}\ and\ \bibinfo
  {author} {\bibfnamefont{F.~H.}\ \bibnamefont{Stillinger}},\ }%
  \bibfield{journal}{%
  \Doi{10.1103/RevModPhys.82.2633}{\bibinfo {journal} {Rev. Mod. Phys.}}\ }%
  \textbf{\bibinfo {volume} {82}},\ \bibinfo {pages} {2633} (\bibinfo {year}
  {2010})%
  \bibAnnoteFile{NoStop}{Torquato10}%
\bibitem{Raitzsch08}%
  \BibitemOpen
  \bibfield{author}{%
  \bibinfo {author} {\bibfnamefont{U.}~\bibnamefont{Raitzsch}}, \bibinfo
  {author} {\bibfnamefont{V.}~\bibnamefont{Bendkowsky}}, \bibinfo {author}
  {\bibfnamefont{R.}~\bibnamefont{Heidemann}}, \bibinfo {author}
  {\bibfnamefont{B.}~\bibnamefont{Butscher}}, \bibinfo {author}
  {\bibfnamefont{R.}~\bibnamefont{L\"{o}w}},\ and\ \bibinfo {author}
  {\bibfnamefont{T.}~\bibnamefont{Pfau}},\ }%
  \bibfield{journal}{%
  \bibinfo {journal} {Phys. Rev. Lett.}\ }%
  \textbf{\bibinfo {volume} {100}},\ \bibinfo {eid} {013002} (\bibinfo {year}
  {2008})%
  \bibAnnoteFile{NoStop}{Raitzsch08}%
\bibitem{Reetz-Lamour08}%
  \BibitemOpen
  \bibfield{author}{%
  \bibinfo {author} {\bibfnamefont{M.}~\bibnamefont{Reetz-Lamour}}, \bibinfo
  {author} {\bibfnamefont{T.}~\bibnamefont{Amthor}}, \bibinfo {author}
  {\bibfnamefont{J.}~\bibnamefont{Deiglmayr}},\ and\ \bibinfo {author}
  {\bibfnamefont{M.}~\bibnamefont{Weidem\"{u}ller}},\ }%
  \bibfield{journal}{%
  \bibinfo {journal} {Phys. Rev. Lett.}\ }%
  \textbf{\bibinfo {volume} {100}},\ \bibinfo {eid} {253001} (\bibinfo {year}
  {2008})%
  \bibAnnoteFile{NoStop}{Reetz-Lamour08}%
\bibitem{Pohl10}%
  \BibitemOpen
  \bibfield{author}{%
  \bibinfo {author} {\bibfnamefont{T.}~\bibnamefont{Pohl}}, \bibinfo {author}
  {\bibfnamefont{E.}~\bibnamefont{Demler}},\ and\ \bibinfo {author}
  {\bibfnamefont{M.~D.}\ \bibnamefont{Lukin}},\ }%
  \bibfield{journal}{%
  \Doi{10.1103/PhysRevLett.104.043002}{\bibinfo {journal} {Phys. Rev. Lett.}}\
  }%
  \textbf{\bibinfo {volume} {104}},\ \bibinfo {pages} {043002} (\bibinfo {year}
  {2010})%
  \bibAnnoteFile{NoStop}{Pohl10}%
\bibitem{Schachenmayer11}%
  \BibitemOpen
  \bibfield{author}{%
  \bibinfo {author} {\bibfnamefont{J.}~\bibnamefont{Schachenmayer}}, \bibinfo
  {author} {\bibfnamefont{I.}~\bibnamefont{Lesanovsky}}, \bibinfo {author}
  {\bibfnamefont{A.}~\bibnamefont{Micheli}},\ and\ \bibinfo {author}
  {\bibfnamefont{A.}~\bibnamefont{Daley}},\ }%
  \bibfield{journal}{%
  \bibinfo {journal} {New J. Phys}\ }%
  \textbf{\bibinfo {volume} {13}},\ \bibinfo {pages} {059503} (\bibinfo {year}
  {2011})%
  \bibAnnoteFile{NoStop}{Schachenmayer11}%
\bibitem{Bijnen11}%
  \BibitemOpen
  \bibfield{author}{%
  \bibinfo {author} {\bibfnamefont{R.~M.~W.}\ \bibnamefont{van Bijnen}},
  \bibinfo {author} {\bibfnamefont{S.}~\bibnamefont{Smit}}, \bibinfo {author}
  {\bibfnamefont{K.~A.~H.}\ \bibnamefont{van Leeuwen}}, \bibinfo {author}
  {\bibfnamefont{E.~J.~D.}\ \bibnamefont{Vredenbregt}},\ and\ \bibinfo {author}
  {\bibfnamefont{S.~J. J. M.~F.}\ \bibnamefont{Kokkelmans}},\ }%
  \bibfield{journal}{%
  \bibinfo {journal} {J. Phys. B}\ }%
  \textbf{\bibinfo {volume} {44}},\ \bibinfo {pages} {184008} (\bibinfo {year}
  {2011})%
  \bibAnnoteFile{NoStop}{Bijnen11}%
\bibitem{Robicheaux05}%
  \BibitemOpen
  \bibfield{author}{%
  \bibinfo {author} {\bibfnamefont{F.}~\bibnamefont{Robicheaux}}\ and\ \bibinfo
  {author} {\bibfnamefont{J.~V.}\ \bibnamefont{Hern\'andez}},\ }%
  \bibfield{journal}{%
  \Doi{10.1103/PhysRevA.72.063403}{\bibinfo {journal} {Phys. Rev. A}}\ }%
  \textbf{\bibinfo {volume} {72}},\ \bibinfo {pages} {063403} (\bibinfo {year}
  {2005})%
  \bibAnnoteFile{NoStop}{Robicheaux05}%
\bibitem{Wuester10}%
  \BibitemOpen
  \bibfield{author}{%
  \bibinfo {author} {\bibfnamefont{S.}~\bibnamefont{W\"uster}}, \bibinfo
  {author} {\bibfnamefont{J.}~\bibnamefont{Stanojevic}}, \bibinfo {author}
  {\bibfnamefont{C.}~\bibnamefont{Ates}}, \bibinfo {author}
  {\bibfnamefont{T.}~\bibnamefont{Pohl}}, \bibinfo {author}
  {\bibfnamefont{P.}~\bibnamefont{Deuar}}, \bibinfo {author}
  {\bibfnamefont{J.~F.}\ \bibnamefont{Corney}},\ and\ \bibinfo {author}
  {\bibfnamefont{J.~M.}\ \bibnamefont{Rost}},\ }%
  \bibfield{journal}{%
  \Doi{10.1103/PhysRevA.81.023406}{\bibinfo {journal} {Phys. Rev. A}}\ }%
  \textbf{\bibinfo {volume} {81}},\ \bibinfo {pages} {023406} (\bibinfo {year}
  {2010})%
  \bibAnnoteFile{NoStop}{Wuester10}%
\bibitem{Schwarzkopf11}%
  \BibitemOpen
  \bibfield{author}{%
  \bibinfo {author} {\bibfnamefont{A.}~\bibnamefont{Schwarzkopf}}, \bibinfo
  {author} {\bibfnamefont{R.~E.}\ \bibnamefont{Sapiro}},\ and\ \bibinfo
  {author} {\bibfnamefont{G.}~\bibnamefont{Raithel}},\ }%
  \bibfield{journal}{%
  \Doi{10.1103/PhysRevLett.107.103001}{\bibinfo {journal} {Phys. Rev. Lett.}}\
  }%
  \textbf{\bibinfo {volume} {107}},\ \bibinfo {pages} {103001} (\bibinfo {year}
  {2011})%
  \bibAnnoteFile{NoStop}{Schwarzkopf11}%
\bibitem{Heidemann07}%
  \BibitemOpen
  \bibfield{author}{%
  \bibinfo {author} {\bibfnamefont{R.}~\bibnamefont{Heidemann}}, \bibinfo
  {author} {\bibfnamefont{U.}~\bibnamefont{Raitzsch}}, \bibinfo {author}
  {\bibfnamefont{V.}~\bibnamefont{Bendkowsky}}, \bibinfo {author}
  {\bibfnamefont{B.}~\bibnamefont{Butscher}}, \bibinfo {author}
  {\bibfnamefont{R.}~\bibnamefont{L\"{o}w}}, \bibinfo {author}
  {\bibfnamefont{L.}~\bibnamefont{Santos}},\ and\ \bibinfo {author}
  {\bibfnamefont{T.}~\bibnamefont{Pfau}},\ }%
  \bibfield{journal}{%
  \bibinfo {journal} {Phys. Rev. Lett.}\ }%
  \textbf{\bibinfo {volume} {99}},\ \bibinfo {eid} {163601} (\bibinfo {year}
  {2007})%
  \bibAnnoteFile{NoStop}{Heidemann07}%
\bibitem{Loew09}%
  \BibitemOpen
  \bibfield{author}{%
  \bibinfo {author} {\bibfnamefont{R.}~\bibnamefont{L\"ow}}, \bibinfo {author}
  {\bibfnamefont{H.}~\bibnamefont{Weimer}}, \bibinfo {author}
  {\bibfnamefont{U.}~\bibnamefont{Krohn}}, \bibinfo {author}
  {\bibfnamefont{R.}~\bibnamefont{Heidemann}}, \bibinfo {author}
  {\bibfnamefont{V.}~\bibnamefont{Bendkowsky}}, \bibinfo {author}
  {\bibfnamefont{B.}~\bibnamefont{Butscher}}, \bibinfo {author}
  {\bibfnamefont{H.~P.}\ \bibnamefont{B\"uchler}},\ and\ \bibinfo {author}
  {\bibfnamefont{T.}~\bibnamefont{Pfau}},\ }%
  \bibfield{journal}{%
  \Doi{10.1103/PhysRevA.80.033422}{\bibinfo {journal} {Phys. Rev. A}}\ }%
  \textbf{\bibinfo {volume} {80}},\ \bibinfo {pages} {033422} (\bibinfo {year}
  {2009})%
  \bibAnnoteFile{NoStop}{Loew09}%
\bibitem{Weimer08}%
  \BibitemOpen
  \bibfield{author}{%
  \bibinfo {author} {\bibfnamefont{H.}~\bibnamefont{Weimer}}, \bibinfo {author}
  {\bibfnamefont{R.}~\bibnamefont{L\"{o}w}}, \bibinfo {author}
  {\bibfnamefont{T.}~\bibnamefont{Pfau}},\ and\ \bibinfo {author}
  {\bibfnamefont{H.~P.}\ \bibnamefont{B\"{u}chler}},\ }%
  \bibfield{journal}{%
  \bibinfo {journal} {Phys. Rev. Lett.}\ }%
  \textbf{\bibinfo {volume} {101}},\ \bibinfo {pages} {250601} (\bibinfo {year}
  {2008})%
  \bibAnnoteFile{NoStop}{Weimer08}%
\bibitem{Olmos09-2}%
  \BibitemOpen
  \bibfield{author}{%
  \bibinfo {author} {\bibfnamefont{B.}~\bibnamefont{Olmos}}, \bibinfo {author}
  {\bibfnamefont{M.}~\bibnamefont{M\"{u}ller}},\ and\ \bibinfo {author}
  {\bibfnamefont{I.}~\bibnamefont{Lesanovsky}},\ }%
  \bibfield{journal}{%
  \bibinfo {journal} {New Journal of Physics}\ }%
  \textbf{\bibinfo {volume} {12}},\ \bibinfo {pages} {013024} (\bibinfo {year}
  {2010})%
  \bibAnnoteFile{NoStop}{Olmos09-2}%
\bibitem{Lesanovsky10}%
  \BibitemOpen
  \bibfield{author}{%
  \bibinfo {author} {\bibfnamefont{I.}~\bibnamefont{Lesanovsky}}, \bibinfo
  {author} {\bibfnamefont{B.}~\bibnamefont{Olmos}},\ and\ \bibinfo {author}
  {\bibfnamefont{J.~P.}\ \bibnamefont{Garrahan}},\ }%
  \bibfield{journal}{%
  \Doi{10.1103/PhysRevLett.105.100603}{\bibinfo {journal} {Phys. Rev. Lett.}}\
  }%
  \textbf{\bibinfo {volume} {105}},\ \bibinfo {pages} {100603} (\bibinfo {year}
  {2010})%
  \bibAnnoteFile{NoStop}{Lesanovsky10}%
\bibitem{Ates11-2}%
  \BibitemOpen
  \bibfield{author}{%
  \bibinfo {author} {\bibfnamefont{C.}~\bibnamefont{Ates}}, \bibinfo {author}
  {\bibfnamefont{J.~P.}\ \bibnamefont{Garrahan}},\ and\ \bibinfo {author}
  {\bibfnamefont{I.}~\bibnamefont{Lesanovsky}},\ }%
  \bibfield{journal}{%
  \bibinfo {journal} {Phys. Rev. Lett. (accepted); preprint},\ \bibinfo {pages}
  {arXiv:1108.0270\,}}%
   (\bibinfo {year} {2011})%
  \bibAnnoteFile{NoStop}{Ates11-2}%
\bibitem{Tonks36}%
  \BibitemOpen
  \bibfield{author}{%
  \bibinfo {author} {\bibfnamefont{L.}~\bibnamefont{Tonks}},\ }%
  \bibfield{journal}{%
  \Doi{10.1103/PhysRev.50.955}{\bibinfo {journal} {Phys. Rev.}}\ }%
  \textbf{\bibinfo {volume} {50}},\ \bibinfo {pages} {955} (\bibinfo {year}
  {1936})%
  \bibAnnoteFile{NoStop}{Tonks36}%
\bibitem{Lesanovsky11}%
  \BibitemOpen
  \bibfield{author}{%
  \bibinfo {author} {\bibfnamefont{I.}~\bibnamefont{Lesanovsky}},\ }%
  \bibfield{journal}{%
  \Doi{10.1103/PhysRevLett.106.025301}{\bibinfo {journal} {Phys. Rev. Lett.}}\
  }%
  \textbf{\bibinfo {volume} {106}},\ \bibinfo {pages} {025301} (\bibinfo {year}
  {2011})%
  \bibAnnoteFile{NoStop}{Lesanovsky11}%
\bibitem{Ji11}%
  \BibitemOpen
  \bibfield{author}{%
  \bibinfo {author} {\bibfnamefont{S.}~\bibnamefont{Ji}}, \bibinfo {author}
  {\bibfnamefont{C.}~\bibnamefont{Ates}},\ and\ \bibinfo {author}
  {\bibfnamefont{I.}~\bibnamefont{Lesanovsky}},\ }%
  \bibfield{journal}{%
  \Doi{10.1103/PhysRevLett.107.060406}{\bibinfo {journal} {Phys. Rev. Lett.}}\
  }%
  \textbf{\bibinfo {volume} {107}},\ \bibinfo {pages} {060406} (\bibinfo {year}
  {2011})%
  \bibAnnoteFile{NoStop}{Ji11}%
\bibitem{Chowdhury00}%
  \BibitemOpen
  \bibfield{author}{%
  \bibinfo {author} {\bibfnamefont{D.}~\bibnamefont{Chowdhury}}\ and\ \bibinfo
  {author} {\bibfnamefont{D.}~\bibnamefont{Stauffer}},\ }%
  \emph{\bibinfo {title} {Principles of Equilibrium Statistical Mechanics}}\
  (\bibinfo {publisher} {Whiley-VCH, Weinheim, Germany},\ \bibinfo {year}
  {2000})%
  \bibAnnoteFile{NoStop}{Chowdhury00}%
\bibitem{Salsburg53}%
  \BibitemOpen
  \bibfield{author}{%
  \bibinfo {author} {\bibfnamefont{Z.~W.}\ \bibnamefont{Salsburg}}, \bibinfo
  {author} {\bibfnamefont{R.~W.}\ \bibnamefont{Zwanzig}},\ and\ \bibinfo
  {author} {\bibfnamefont{J.~G.}\ \bibnamefont{Kirkwood}},\ }%
  \bibfield{journal}{%
  \bibinfo {journal} {J. Chem. Phys.}\ }%
  \textbf{\bibinfo {volume} {21}},\ \bibinfo {pages} {1098} (\bibinfo {year}
  {1953})%
  \bibAnnoteFile{NoStop}{Salsburg53}%
\bibitem{Viteau11}%
  \BibitemOpen
  \bibfield{author}{%
  \bibinfo {author} {\bibfnamefont{M.}~\bibnamefont{Viteau}}, \bibinfo {author}
  {\bibfnamefont{M.~G.}\ \bibnamefont{Bason}}, \bibinfo {author}
  {\bibfnamefont{J.}~\bibnamefont{Radogostowicz}}, \bibinfo {author}
  {\bibfnamefont{N.}~\bibnamefont{Malossi}}, \bibinfo {author}
  {\bibfnamefont{D.}~\bibnamefont{Ciampini}}, \bibinfo {author}
  {\bibfnamefont{O.}~\bibnamefont{Morsch}},\ and\ \bibinfo {author}
  {\bibfnamefont{E.}~\bibnamefont{Arimondo}},\ }%
  \bibfield{journal}{%
  \Doi{10.1103/PhysRevLett.107.060402}{\bibinfo {journal} {Phys. Rev. Lett.}}\
  }%
  \textbf{\bibinfo {volume} {107}},\ \bibinfo {pages} {060402} (\bibinfo {year}
  {2011})%
  \bibAnnoteFile{NoStop}{Viteau11}%
\bibitem{Buechler07}%
  \BibitemOpen
  \bibfield{author}{%
  \bibinfo {author} {\bibfnamefont{H.~P.}\ \bibnamefont{B\"uchler}}, \bibinfo
  {author} {\bibfnamefont{E.}~\bibnamefont{Demler}}, \bibinfo {author}
  {\bibfnamefont{M.}~\bibnamefont{Lukin}}, \bibinfo {author}
  {\bibfnamefont{A.}~\bibnamefont{Micheli}}, \bibinfo {author}
  {\bibfnamefont{N.}~\bibnamefont{Prokof'ev}}, \bibinfo {author}
  {\bibfnamefont{G.}~\bibnamefont{Pupillo}},\ and\ \bibinfo {author}
  {\bibfnamefont{P.}~\bibnamefont{Zoller}},\ }%
  \bibfield{journal}{%
  \Doi{10.1103/PhysRevLett.98.060404}{\bibinfo {journal} {Phys. Rev. Lett.}}\
  }%
  \textbf{\bibinfo {volume} {98}},\ \bibinfo {pages} {060404} (\bibinfo {year}
  {2007})%
  \bibAnnoteFile{NoStop}{Buechler07}%
\bibitem{Micheli07}%
  \BibitemOpen
  \bibfield{author}{%
  \bibinfo {author} {\bibfnamefont{A.}~\bibnamefont{Micheli}}, \bibinfo
  {author} {\bibfnamefont{G.}~\bibnamefont{Pupillo}}, \bibinfo {author}
  {\bibfnamefont{H.~P.}\ \bibnamefont{B\"{u}chler}},\ and\ \bibinfo {author}
  {\bibfnamefont{P.}~\bibnamefont{Zoller}},\ }%
  \bibfield{journal}{%
  \bibinfo {journal} {Phys. Rev. A}\ }%
  \textbf{\bibinfo {volume} {76}},\ \bibinfo {pages} {043604} (\bibinfo {year}
  {2007})%
  \bibAnnoteFile{NoStop}{Micheli07}%
\bibitem{Gaerttner12}%
  \BibitemOpen
  \bibfield{author}{%
  \bibinfo {author} {\bibfnamefont{M.}~\bibnamefont{G\"arttner}}\ and\ \bibinfo
  {author} {\bibfnamefont{J.}~\bibnamefont{Evers}},\ }%
  \bibinfo {journal} {private communation}%
  \bibAnnoteFile{NoStop}{Gaerttner12}%
\bibitem{Olmos11}%
  \BibitemOpen
\bibfield{journal}{%
    }%
  \bibfield{author}{%
  \bibinfo {author} {\bibfnamefont{B.}~\bibnamefont{Olmos}}, \bibinfo {author}
  {\bibfnamefont{W.}~\bibnamefont{Li}}, \bibinfo {author}
  {\bibfnamefont{S.}~\bibnamefont{Hofferberth}},\ and\ \bibinfo {author}
  {\bibfnamefont{I.}~\bibnamefont{Lesanovsky}},\ }%
  \bibfield{journal}{%
  \Doi{10.1103/PhysRevA.84.041607}{\bibinfo {journal} {Phys. Rev. A}}\ }%
  \textbf{\bibinfo {volume} {84}},\ \bibinfo {pages} {041607} (\bibinfo {year}
  {2011})%
  \bibAnnoteFile{NoStop}{Olmos11}%
\bibitem{Gunter12}%
  \BibitemOpen
  \bibfield{author}{%
  \bibinfo {author} {\bibfnamefont{G.}~\bibnamefont{G\"unter}}, \bibinfo
  {author} {\bibfnamefont{M.}~\bibnamefont{{Robert-de-Saint-Vincent}}},
  \bibinfo {author} {\bibfnamefont{H.}~\bibnamefont{Schempp}}, \bibinfo
  {author} {\bibfnamefont{C.~S.}\ \bibnamefont{Hofmann}}, \bibinfo {author}
  {\bibfnamefont{S.}~\bibnamefont{Whitlock}},\ and\ \bibinfo {author}
  {\bibfnamefont{M.}~\bibnamefont{Weidem\"uller}},\ }%
  \bibfield{journal}{%
  \Doi{10.1103/PhysRevLett.108.013002}{\bibinfo {journal} {Phys. Rev. Lett.}}\
  }%
  \textbf{\bibinfo {volume} {108}},\ \bibinfo {pages} {013002} (\bibinfo {year}
  {2012})%
  \bibAnnoteFile{NoStop}{Gunter12}%
\end{thebibliography}

%Merlin.mbs v4.21 2009-07-09.
%

\end{document}